# Generalized counter-rotating oscillators: Mixed synchronization

Sourav K. Bhowmick[1,2], Bidesh K. Bera[3], Dibakar Ghosh[3*]

[1]*Department of Electronics, Asutosh College, Kolkata 700 026, India.*
[2]*Central Instrumentation, CSIR-Indian Institute of Chemical Biology, Jadavpur, Kolkata 700032, India.*
[3]*Physics and Applied Mathematics Unit, Indian Statistical Institute, Kolkata-700108, India.*

*Corresponding author. Email adess: dibakar@isical.ac.in (D. Ghosh).

In this paper, we report mixed synchronization between two counter rotating chaotic oscillators. We describe a procedure how to obtain a counter rotating oscillator for generalized oscillators. We elaborate the method with numerical examples of the Sprott system, Pikovsky-Rabinovich (PR) circuit model. Noise-induced mixed synchronization is also reported in PR circuit model. The physical realization of mixed synchronization in an electronic circuit of two counters-rotating Sprott systems also shown.

## I.   Introduction

Counter rotating oscillators, where two oscillators rotate in opposite direction in their phase space, is the recent interest in dynamical systems. The counter rotating oscillators exist in natural systems in the field of fluid dynamics [1], biological systems [2] and also physical systems[3-5]. Czolczynski et al.[3] observed the dynamics of system of $n$ pendula mounted on the movable bean where one pendulum rotates counterclockwise direction with positive angular velocity and the remaining pendula rotate in clockwise direction with negative angular velocity. Recently, Prasad [6] observed the counter rotating oscillators and dynamical behaviour when they are coupled. But the general mathematical description of how to obtain counter rotating oscillators from a given dynamical system is proposed by Bhowmick *et. al.* [7]. Although there is another way to make counter rotation of a given dynamical systems by replacing any state variable by same state variable with negative sign, but creation of counter rotation is not guaranteed by this replacement method. When two counter-rotating oscillators are interacting with each other, a mixed type of synchronization (MS) is observed, i.e. some of the states variables are in complete synchronize (CS) state and other variables are in anti-synchronize (AS) state. MS is also obtained between two identical and mismatched oscillators in delayed [8] and non-delayed [9] systems. Previously, existence of MS state is observed in co-rotating Lorenz oscillator with a specific scalar coupling due to inherent axial symmetry [9] but in general, MS is not observed in all co-rotating oscillators. A design of coupling for engineering a MS state in chaotic systems was proposed [10] recently. In recent study [7], a general mathematical description is given how to obtain a counter-rotating oscillator from a given dynamical systems where only one pair of conjugate elements in the linear matrix are non-zero. But the general mechanism of how to obtain a counter rotating oscillator from a given dynamical systems where more than one conjugate elements are non-zero in the linear matrix of order $3 \times 3$ and investigation of collective behaviour between two couple counter-rotating oscillators is still missing. The question of counter-rotation in a dynamical system was first discussed by Tabor [11] in relation to rotation of a two-dimensional (2D) limit cycle system near a Hopf bifurcation point. The concept of the counter-rotating oscillators was extended [6] to investigate their collective behaviours and further applied to chaotic system as well.

Many situations in nature arises in which two or more oscillators that interact with each other through a common medium or common external noise. Common noise is also great relevance to biological systems, neural networks and laser science [12]. Noise is ubiquitous in nature and manmade systems can play as a constructive as well as destructive role. The constructive role of noise as enhancement of phase synchronization is established [13] recently in models and also in experiments of chemical oscillators [14], although, stochastic resonance [15] as a constructive role of noise is well known for more than two decades. Also two or more oscillators are coupled via a common medium are synchronized when they are not interacting with each other [16]. External noise can induce synchronized between two or more uncoupled oscillators. The destructive effect of noise is mostly studied earlier as a test for robustness of synchronization since noise is always present in any experiment or in nature. Numerical and experimental studies on noise-induce-synchronization (NIS) have been found in many physical systems such as laser, electronic circuits, biological systems and neuronal networks [17]. In neural systems, NIS was first observed between a pair of uncoupled sensory neurons [18]. In the previous study of NIS, CS and PS are observed where all the state variables of two uncoupled systems are in same phase. But noise-induced mixed synchronization in chaotic oscillators is not reported so far, to our best knowledge.

In this paper we extend the study of MS in general oscillators where more than one pair of conjugate elements of the linear matrix is non-zero. A particular type of MS is observed using different choice of scalar coupling between two identical counter-rotating oscillators. Noise-induced MS is also possible in counter-rotating oscillators. We elaborate the method of MS using numerical simulation on Sprott system [19], Pikovsky-Rabinovich (PR) circuit model [20]. For limit cycle oscillators, the general mechanism to create a counter-rotation oscillator is explained in details but MS in couple counter-rotating limit cycle oscillators is not shown in this paper. Finally, we experimentally demonstrate the practicality of the method using two coupled electronic circuit of Sprott system.

The plan of the paper is as follows. In sec. II, we discuss how to create a counter rotating limit cycle oscillator and numerically shown using Landou-Stuart [21] and Van der Pol [22] oscillators. Then we extend the theory for chaotic oscillators where more than one pair of conjugate elements in the linear matrix of order $3 \times 3$ is non-zero. In sec. III, the emergence of MS is such coupled counter-rotating chaotic oscillators are discussed. Noise-induced MS in coupled counter-rotating PR model is discussed in sec.IV. Experimental observation using electronic circuit of Sprott system is discussed in sec. V. Finally, in sec. VI, we summarize our results.

## II. Generalized counter-rotating oscillators: General theory and numerical simulations

We consider a two-dimension (2D) limit cycle oscillator in general form as

$$\dot{x} = a_{11}x + a_{12}y + h_1(x, y)$$
$$\dot{y} = a_{21}x + a_{22}y + h_2(x, y)$$
(1)

where $f(x, y)$ and $g(x, y)$ are non-linear functions. The above system can be written as

$$\dot{X} = AX + H(X)$$
(2)

where $X = (x, y)^T$ is a state variable, $T$ denotes transpose of matrix, $A = \begin{pmatrix} a_{11} & a_{12} \\ a_{21} & a_{22} \end{pmatrix}$ is constant linear matrix and $H(X) = [h_1(x, y), h_2(x, y)]^T$ is nonlinear function. If the system (2) is hyperbolic then using Lyapunov stability property [23] (unstable, stable, uniformly stable, asymptotically stable, uniformly and asymptotically stable) the solution of system (2) is same as that the solution of the homogeneous equation

$$\dot{Y} = AY \text{ where } Y = (\xi, \eta)^T \tag{3}$$

The eigenvalues of the linear matrix $A$ are

$$\lambda_{1,2} = \frac{(a_{11} + a_{22}) \pm \sqrt{(a_{11} + a_{22})^2 - 4(a_{11}a_{22} - a_{12}a_{21})}}{2} \tag{4}$$

The direction of rotation is only applicable when the equilibrium point is a centre or spiral or focus. The condition for that is

$$a_{12}a_{21} < -\frac{(a_{11} - a_{22})^2}{4} \tag{5}$$

To obtain the direction of rotation, it is sufficient to use the linear equation (3) (or the original equation (2) may be used), putting $\xi > 0, \eta = 0$ then $\dot{\eta} = a_{21}\xi > 0$ (or $< 0$) according to $a_{21} > 0$ (or $< 0$), indicating that the rotation is in counterclockwise (or clockwise) direction. On the other hand, by putting $\eta > 0, \xi = 0$ then obtain $\dot{\xi} = a_{12}\eta$. Similar to above, the rotation is in counterclockwise (or clockwise) direction according to $a_{12} > 0$ (or $a_{12} < 0$). For counter-rotating rotation, we have to change the sign of either $a_{21}$ or $a_{12}$. But from condition (5), it is easy to see that the sign of both $a_{12}$ and $a_{21}$, which are the conjugate elements of the linear constant matrix $A$, are to be changed. The counter-rotating oscillator is possible if conjugate elements $a_{12}$ and $a_{21}$ both exist and replace them by its negative values. Next we consider examples to examine that change of sign of conjugate elements of the linear matrix $A$ create a counter-rotating oscillators or not. We use two examples, a Landau-Stuart [21] and Van der Pol [22] oscillator. As a first example, we analyze a model of Landau-Stuart oscillators [21] given by the equation of motion

$$\dot{z} = (1 + i\omega - |z|^2)z \tag{6}$$

with $\omega$ is the natural frequency and $z(t) = x(t) + iy(t)$ The dynamical equation in Cartesian form is

$$\begin{aligned} \dot{x} &= [1 - (x^2 + y^2)]x - \omega y \\ \dot{y} &= [1 - (x^2 + y^2)]y + \omega x \end{aligned} \tag{7}$$

The above system has only one fixed point (0, 0) and which is unstable spiral. The linear matrix $A$ for counter clockwise rotation of system (7) and a clockwise rotation by changes it by $A'$ are given by

$$A = \begin{pmatrix} 1 & -\omega \\ \omega & 1 \end{pmatrix} \text{ and } A' = \begin{pmatrix} 1 & \omega \\ -\omega & 1 \end{pmatrix} \tag{8}$$

Here the conjugate elements $a_{12}$, $a_{21}$ are present in the linear matrix and after changing by its negative value, we obtain the linear matrix $A'$. The counter-rotating limit cycle attractors of system (7) are shown in Fig. 1(a) and (b) for $\omega = 2.0$. Next we consider Van der Pol oscillator [22] as a second example in the form

$$\dot{x} = \omega y$$
$$\dot{y} = b(1-x^2)y - \omega x \qquad (9)$$

The counter-rotating limit cycles of system (9) for $\omega = 1.0$ and $\omega = -1.0$ are shown in Fig. 1(c) and (d).

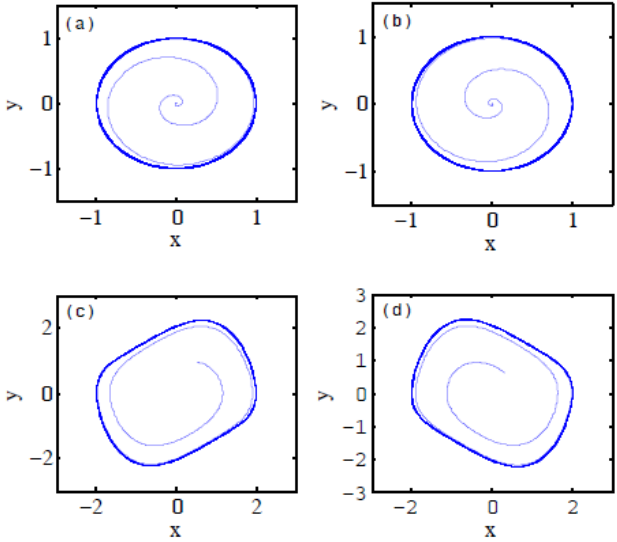

Figure 1: Limit cycles attractors of Landau-Stuart oscillator with linear matrix (a) $A$ and (b) $A'$. counter-rotating phase portrait of Van der Pol oscillators for (c) ω=1 and (d) ω=-1 with parameter b=2.0.

The direction of rotation, clockwise or counter-clockwise of a three-dimensional (3D) dynamical flow is explained as the direction of rotation in 2D coordinate with respect to a third coordinate. Thus we summarize the results for 3D dynamical systems as follows:

**Proposition-I:** A dynamical system can be expressed by,

$$\dot{x} = Ax + f(x) + C \qquad (10)$$

where $x \in R^n$ is the state vector, $A$ is $n \times n$ constant linear matrix, $C$ is a $n \times 1$ constant matrix and $f: R^n \to R^n$ contains the nonlinear function. For 3D systems, $A=(a_{ij})_{3\times 3}$, we obtain a general procedure for deriving counter rotating attractor as stated in the following steps:

i) at least one pair of non-zero element $a_{ij}, a_{ji}, i \neq j$ of the matrix A must exist, where $i, j=1,2,3$.
ii) replace that nonzero $a_{ij}$ by $-a_{ij} (i \neq j)$ and vice-versa.

This rotation is controllable by manipulating the linear matrix as proposed above and rest of the parts of system (10) is unchanged. We apply the above technique for counter rotation on chaotic systems where more than one pair of conjugate elements in the linear matrix is present. The details explanation of this Proposition-I where the one pair of conjugate elements is non-zero are well explained in Ref.[7].

For numerical simulations we consider chaotic oscillators, namely, Sprott system [19] and Pikovsky-Rabinovich circuit model [20]. As a first example, we consider Sprott systems as follows

$$\dot{x}_1 = y_1$$
$$\dot{y}_1 = x_1 - z_1$$
$$\dot{z}_1 = x_1 + x_1 z_1 + a y_1 \qquad (11)$$

which is chaotic for $a = 2.7$. The linear matrix of the Sprott system is

$$A = \begin{bmatrix} 0 & 1 & 0 \\ 1 & 0 & -1 \\ 1 & a & 0 \end{bmatrix}$$

In the linear matrix A, two non-zero conjugate pairs $(a_{12}, a_{21})$ and $(a_{23}, a_{32})$ are exist. So according to the rule stated above (Proposition-I), all conjugate pair elements are replaced by its negative sign. In ref [7], all examples are taken where only one pair of conjugate element is non-zero in the linear matrix. The new linear matrix becomes

$$A' = \begin{bmatrix} 0 & -1 & 0 \\ -1 & 0 & 1 \\ 1 & -a & 0 \end{bmatrix}$$

That produces a counter-rotating Sprott system. Finally counter-rotating Sprott system becomes

$$\dot{x}_2 = -y_2$$
$$\dot{y}_2 = -x_2 + z_2 \qquad (12)$$
$$\dot{z}_2 = x_2 + x_2 z_2 - a y_2$$

In Fig. 2, the chaotic attractors are shown for Sprott systems (11) and (12). It is cleared that only changing the sign of the conjugate pair of elements in the linear matrix, counter-rotation in the trajectories of the dynamical systems are induced. Fig. 2(a) and 2(b) are counter-rotating in $x - y$ plane and Fig. 2(c) and 2(d) are counter rotating in $y - z$ plane. If the parameter $a$ is changed, the two largest Lyapunov exponents of the individual oscillators are almost identical [6] shown in Fig. 3 and only the position of the phase space of two attractors for counter-rotating oscillators are changed.

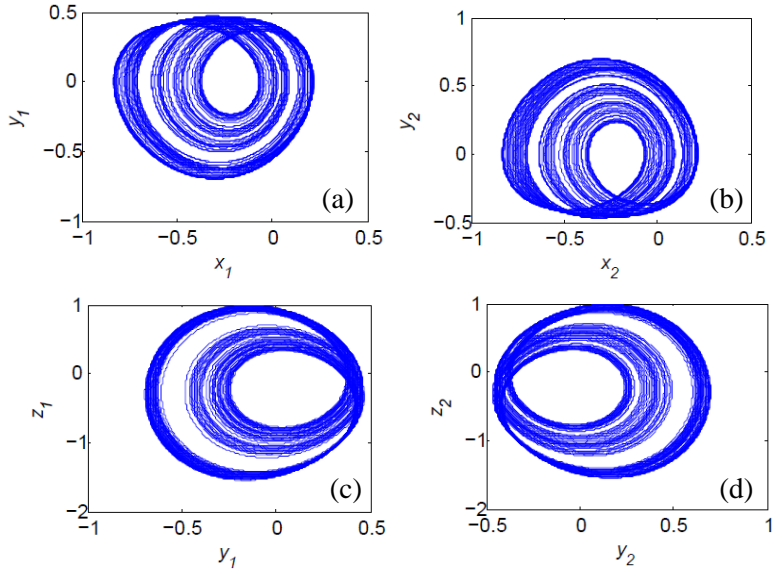

Figure 2: Counter-rotating chaotic attractors of Sprott systems (11) and (12): a) $x_1$-$y_1$ plane (b) $x_2$-$y_2$ plane (c) $y_1$-$z_1$ plane and (d) $y_2$-$z_2$ plane.

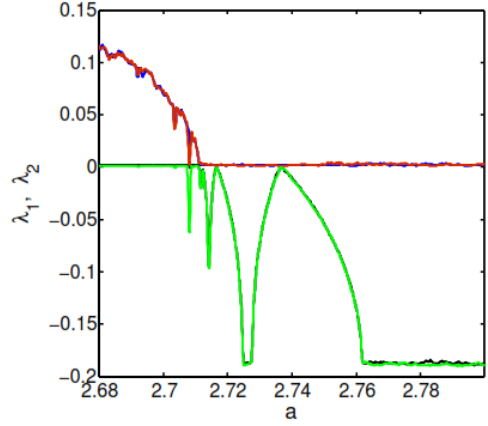

Figure 3: Variation of first two Lyapunov exponent by varying the parameter *a* of the Sprott system (11) and its counter-rotating oscillator (12). The first two Lyapunov exponents of (11) are represented by blue and green color solid lines. Red and black color lines are represented the first two Lyapunov exponents for counter-rotating oscillator (12).

### III: Mixed-synchronization in coupled counter-rotating chaotic systems

Mixed synchronization can be easily achieved by diffusion coupling in one variable in plane of rotation. For Sprott system the plane of rotations are *x-y* and *y-z*. But make sure that to get mixed synchronization, one must be coupled any one of the variables, in which plane the counter rotation is achieved. The complete synchronization is observed in that variable which is coupled, the other variable in the plane of rotation is anti-synchronization state [7]. The counter-rotating coupled Sprott system as defined by,

$$\dot{x}_i = \omega_i y_i + \varepsilon(x_j - x_i)$$
$$\dot{y}_i = \omega_i x_i - \omega_i z_i \qquad (13)$$
$$\dot{z}_i = x_i + x_i z_i + a\omega_i y_i + \varepsilon(z_j - z_i)$$

where $\omega_1 = 1.0, \omega_2 = -1.0$, *a*=2.7, $i,j = 1,2, i \neq j$ and $\epsilon$ is coupling strength.

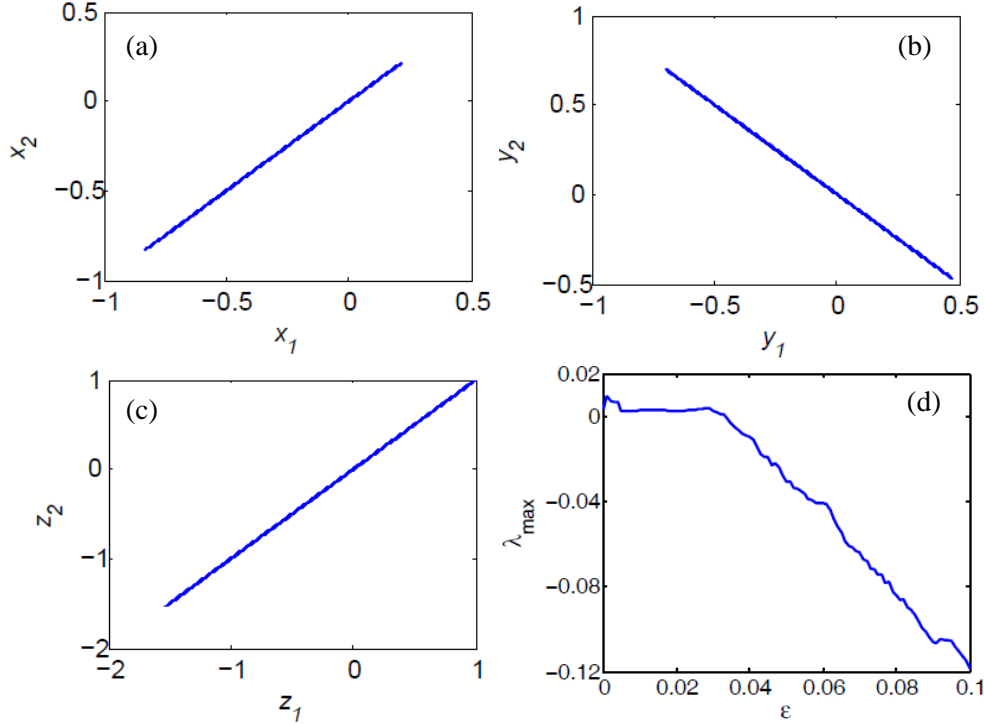

Figure 4: Numerically simulated result of synchronization manifold in coupled counter-rotating Sprott systems (13) : (a) $x_1$-$x_2$ in CS (b) $y_1$-$y_2$ in AS and (c) $z_1$-$z_2$ in CS. (d) Variation of maximum transverse Lyapunov exponent $\lambda_{max}$ with coupling strength ε.

For a value of coupling ε=0.04, MS is observed by simple diffusive coupling and mixed-synchronization manifold are shown in Fig. 4. The $x_1$-$x_2$ and $z_1$-$z_2$ in CS as they are directly coupled and $y_1$-$y_2$ in AS as they are another variable in plane of rotation for both cases. In the previous work [7] the plane of rotation is one since one pair of conjugate element in the linear matrix A non-zero. So for that case synchronization manifold can predict only two variables in the plane of rotation and the third variable is either CS or AS, that will depends on the system. But in this type of system where two pair of conjugate elements in the linear matrix A are non-zero, we can also predict the synchronization behaviour (CS or AS) of the third variable also. There are many possible ways of diffusive coupling, e.g. (i) if we coupled two counter-rotating Sprott systems through $x$-variable only, $x_1$-$x_2$ is in CS and the other variable $y_1$-$y_2$ in x-y plane of rotation is in AS state and $z_1$-$z_2$ will be in CS state since z-variable is another variable in $y-z$ plane, (ii) if we coupled through y-variable only, $y_1 - y_2$ is in CS state, $x_1 - x_2$ and $z_1 - z_2$ are in AS state and so on. MS state is also possible using unidirectional coupling through either $x$ −variable or $y$ −variable or $z$ −variable in place of birectional coupling.

The existence of MS scenario in coupled counter-rotating system can also verify by the calculation of transversal Lyapunov exponents. A necessary condition for the synchronized regime to be stable is that the maximum transversal Lyapunov exponent $\lambda_{max}$ be negative. The linear stability of the MS is determined by the variational equation

$$\begin{aligned}
\dot{e}_1 &= -2\varepsilon\, e_1 + e_2 \\
\dot{e}_2 &= e_1 - e_3 \\
\dot{e}_3 &= (1+z_2)e_1 + ae_2 + (x_1 - 2\varepsilon)e_3
\end{aligned} \qquad (14)$$

where $e_1 = x_1 - x_2$, $e_2 = y_1 + y_2$ and $e_3 = z_1 - z_2$ are the transversal deviation from the synchronization manifold. The variation of this $\lambda_{max}$ with coupling strength $\epsilon$ helps to identify the onset of MS regime. In Fig. 4(d) we plot the $\lambda_{max}$ for coupled Sprott system for various coupling strength $\epsilon$.

As a second example, we consider coupled counter-rotating PR model [20] as follows:

$$\dot{x}_i = \omega_i y_i - \omega_i \beta z_i$$
$$\dot{y}_i = -\omega_i x_i + 2\gamma y_i + \alpha z_i + \varepsilon(y_j - y_i)$$
$$\dot{z}_1 = (\omega_i x_1 - z_1^3 + z_1)/\mu + \varepsilon(z_j - z)$$

(15)

which is chaotic for β=0.66, γ=0.201, α=0.165, μ=0.047, $\omega_1$=1, $\omega_2$=-1, $i \neq j, i, j = 1, 2$ and $\epsilon$ is coupling constant. The linear matrices of rotating and counter-rotating system's are

$$A = \begin{bmatrix} 0 & 1 & -\beta \\ -1 & 2\gamma & \alpha \\ 1/\mu & 0 & 1/\mu \end{bmatrix}, \text{ and } A' = \begin{bmatrix} 0 & -1 & \beta \\ 1 & 2\gamma & \alpha \\ -1/\mu & 0 & 1/\mu \end{bmatrix}$$

The counter rotating attractors of the PR model is shown in Fig.5.

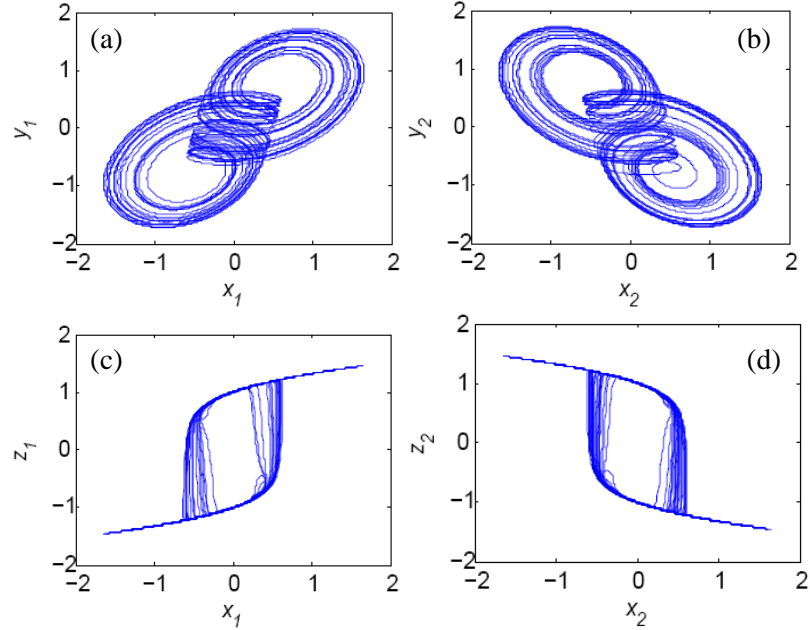

Figure 5: Counter-rotating chaotic attractor of PR model in *x-y* plane with linear matrix (a) *A* and (b) *A'*, in *x-z* plane with linear matrix (c) *A* and (d) *A'*.

Complete synchronization for *y* and *z* variables and anti synchronization for *x* variable can be achieved because both system is coupled via *y* and *z* variables and plane of rotations are in *x-y* and *x-z*. Fig. 6(a) shows the numerical results that confirm AS between $(x_1, x_2)$, Fig. 6(b) shows CS between $(y_1, y_2)$ and Fig. 6(c) shows CS between $(z_1, z_2)$.

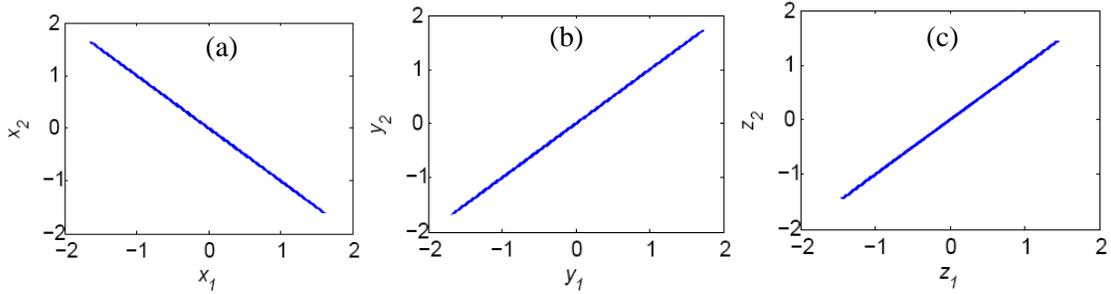

Figure 6: Numerical results of the mixed-synchronization manifold of PR model: (a) $x_1$-$x_2$ in AS, (b) $y_1$-$y_2$ in CS and (c) $z_1$-$z_2$ in CS for $\varepsilon=0.2$.

## IV. Noise-induced mixed synchronization

In this section we discuss noise-induced mixed-synchronization in counter-rotating coupled oscillators. For numerical simulation, we consider counter-rotating Pikovsky-Rabinovich [20] circuit model with common white noise as,

$$\begin{aligned}
\dot{x}_i &= \omega_i y_i - \omega_i \beta z_i \\
\dot{y}_i &= -\omega_i x_i + 2\gamma y_i + \alpha z_i \\
\dot{z}_i &= (\omega_i x_i - z_i^3 + z_i)/\mu + \xi(t)
\end{aligned} \qquad (16)$$

where $\beta=0.66$, $\gamma=0.201$, $\alpha=0.1665$, $\mu=0.047$ and $\omega_1=1$ $\omega_2=-1$. The noise $\xi(t)$ is Gaussian white noise with the properties $<\xi(t)>=0$ and $\langle\xi(t)\xi'(t)\rangle = 2D\delta(t-t')$ where $D>0$ is the noise intensity, $\delta$ is the Dirac delta function, and $<\ldots>$ denotes averaging over the realizations of $\xi(t)$. The equations (16) were integrated using the stochastic Euler method with a time-step of $\delta(t)=0.001$. If the noise intensity $D$ is increases, the MS scenario emerges as shown in figure 7(a)-(c). They confirm AS in $x_1$-$x_2$ pair (Fig. 7(a)), CS in $y_1$-$y_2$ pair (Fig. 7(b)) and CS in $z_1$-$z_2$ pair (Fig. 7(c)). As shown that PR model has plane of rotations in $x$-$y$ and $x$-$z$ plane, we introduced noise in $z$-equation, so $z$-variable is in CS, $x$-variable is in AS and $y$-variable is in CS state. According to the criteria of Pecora and Carroll [24], a negative value of maximum transverse Lyapunov exponent (MTLE) is the necessary condition for noise-induced synchronization. The transverse Lyapunov exponent for the MS is obtained using the variational equation of counter-rotating PR model (16) as

$$\begin{aligned}
\dot{e}_1 &= e_2 - \beta e_3 \\
\dot{e}_2 &= -e_1 + 2\gamma e_2 + \alpha e_3 \\
\dot{e}_3 &= \frac{1}{\mu}e_1 + \frac{1 - z_1^2 - z_1 z_2 - z_2^2}{\mu} e_3
\end{aligned} \qquad (17)$$

where $e_1 = x_1 + x_2, e_2 = y_1 - y_2, e_3 = z_1 - z_2$ are the transverse deviations from the synchronization manifold. Note that, variational equation (17) does not depend on noise explicitly. Here the presence of noise $\xi(t)$ in equation (16) affects the time evolutions of $z_1$, $z_2$ and influence of noise in variational equation (17) occurs via $z_1$ and $z_2$ The MTLE (solid blue line) and average mixed synchronization error

$|X-Y| = < \sqrt{(x_1+x_2)^2 + (y_1-y_2)^2 + (z_1-z_2)^2} >$ (red dashed line) versus noise intensity $D$ is shown in figure 7(d). It is clear from that mixed synchronization error decreases with increasing of noise-intensity $D$. The critical noise intensity is $D^* \approx 2.91$. At this critical point, mixed synchronization error is zero and MTLE is negative which indicated that two counter-rotating PR models are completely mixed synchronized due to their common white noise. The counter-rotating attractors in $x$-$y$ plane are shown in figure 7(e) and 7(f) at the synchronized position for $D=2.92$. In the next section we will discuss about the experimental evidences of MS scenario using electronic circuit.

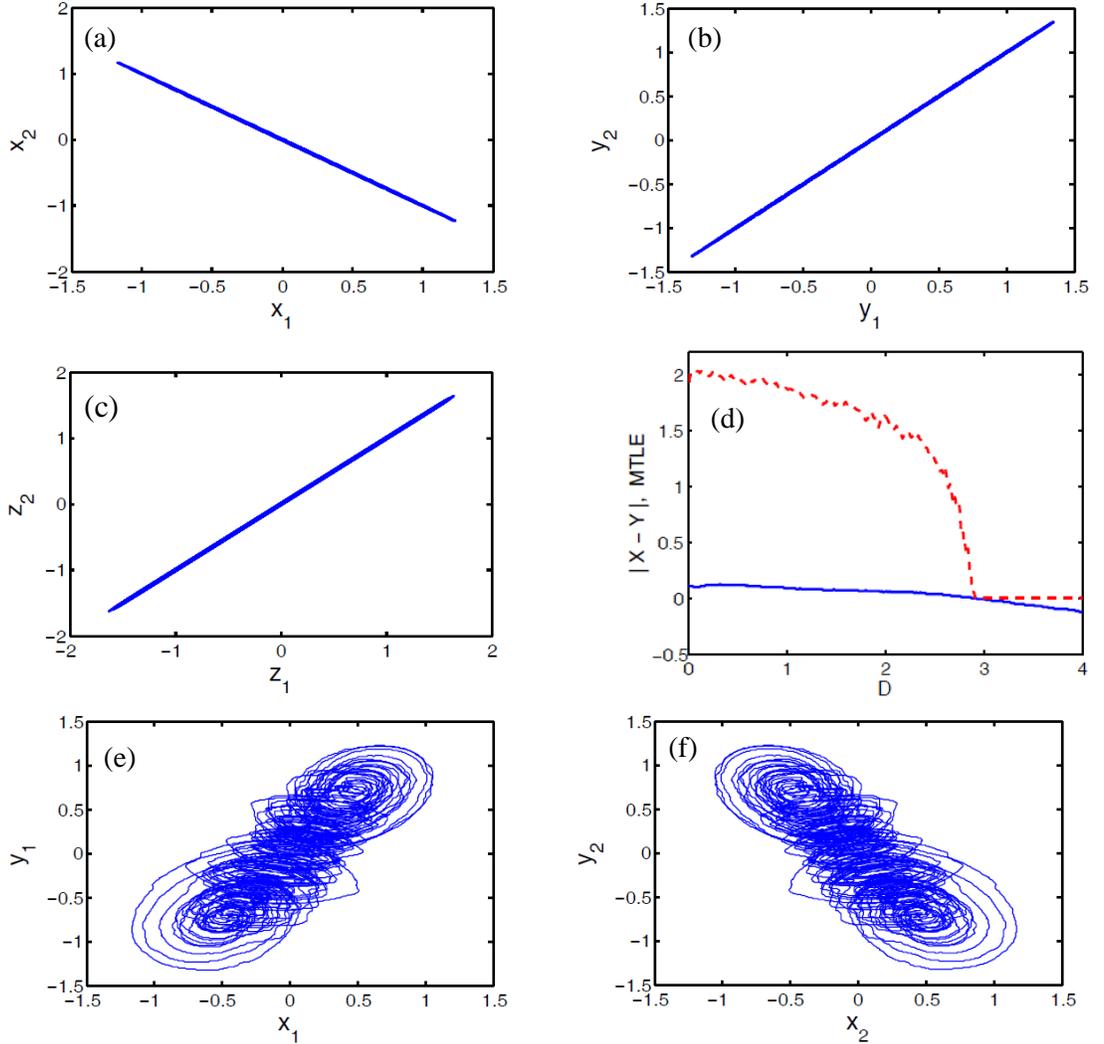

Figure 7: Mixed synchronization manifold (a) $x_1$-$x_2$ is AS, (b) $y_1$-$y_2$ is CS, (c) $z_1$-$z_2$ is CS, (d) Variation of maximum transverse Lyapunov exponent and average mixed synchronization error with noise intensity $D$, chaotic attractors at synchronized state for $D=2.92$ in (e) $x_1$-$y_1$ plane, (f) $x_2$-$y_2$ plane.

## V. Experimental observations

We experimentally verify the MS scenario in counter-rotating oscillators using electronic circuits. In Fig. (8) it represents the Sprott system defined by the equations (11) and (12). The circuit diagram of the Counter rotation of equation (11) is given in Fig. 8(a). The circuit components U1-U3 is integrator U4-U5 is unit gain inverting amplifier designed by operational amplifier (µA741). U6 is multiplier (AD633) which multiplies two input signal to generate nonlinear part of the system. In Fig. 8(b) U7-U9 are integrator U10-U11 in inverting amplifier and U12 is multiplier performs similarly as in Fig. 8(a). The output point of the U1-U3 represents the variables $x_1$, $y_1$ and $z_1$ in form of voltage and similar for U7-U9 represents $x_2$, $y_2$ and $z_2$. The oscilloscope picture of the counter rotating oscillator is shown in Fig. (9) where $x_1$ vs. $y_1$ is plotted in (a), $x_2$ vs. $y_2$ in (b) $y_1$ vs. $z_1$ in (c) and $y_2$ vs. $z_2$ in (d). It is very easy to observe the motion of the attractor is opposite in sense. For getting MS we coupled the counter rotating circuit by simple resistance as shown Fig. 8(c). Here the coupling strength varies inversely with coupling resistance. For a coupling R=15 ohm the both systems are synchronized and shows MS. The experimental synchronization manifolds are shown in Fig. 9(e-g). Here $x_1$-$x_2$ is in CS, $y_1$-$y_2$ in AS and $z_1$-$z_2$ in CS and which is similar to numerical results in Fig. 4(a-c).

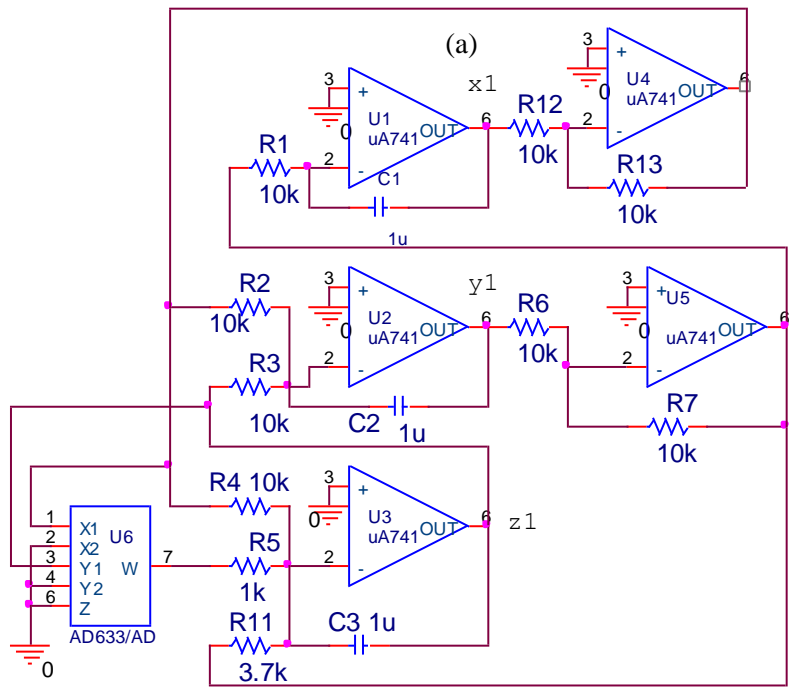

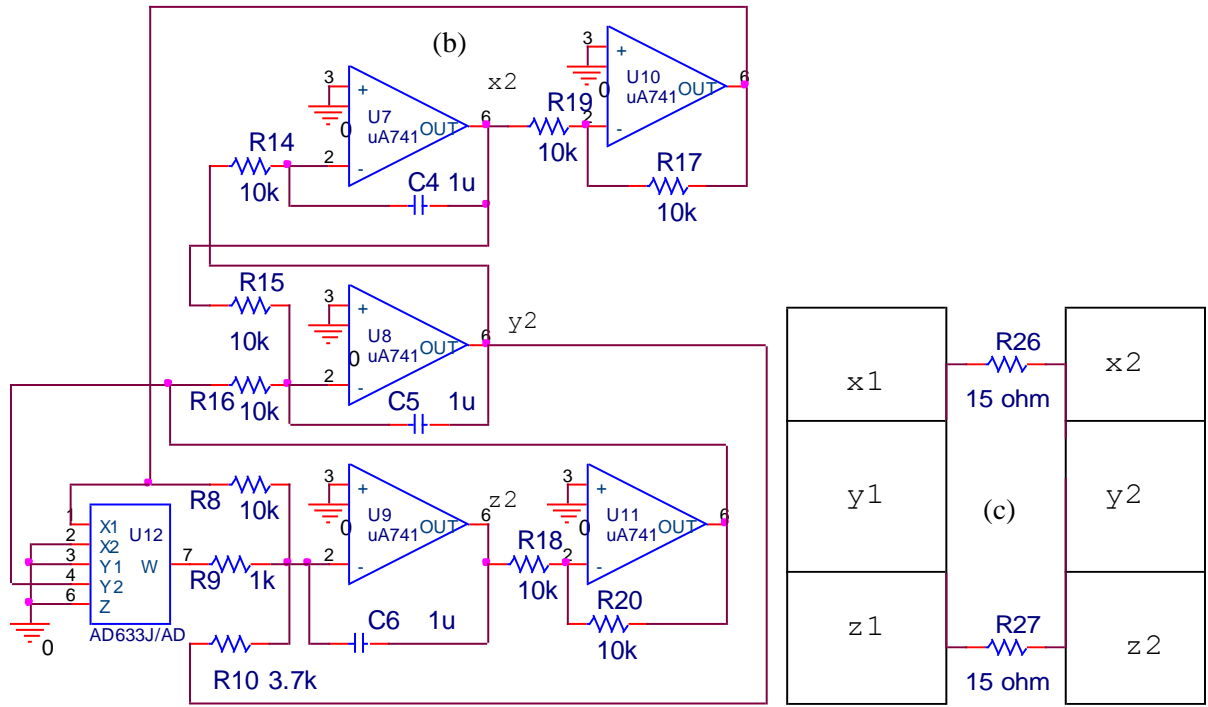

Figure 8: (a) and (b) are the electronic diagram of counter rotating Sprott oscillators. The coupling scheme (c) for getting MS.

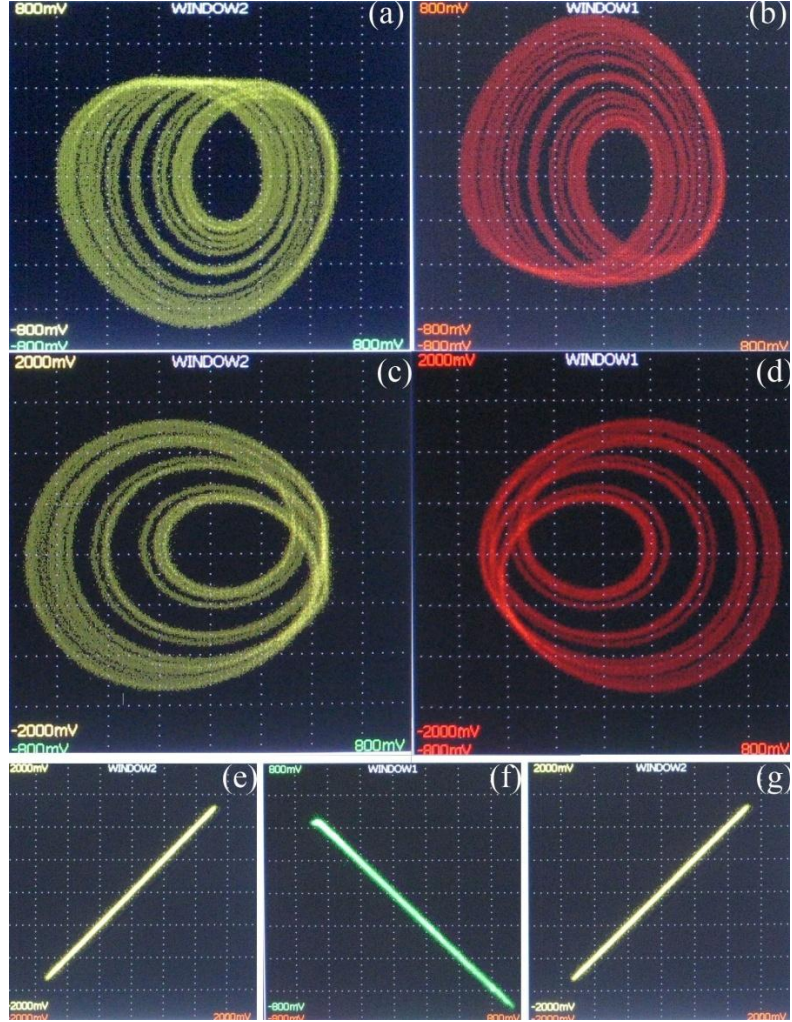

Figure9: Oscilloscope picture of two counter rotating oscillator of Sprott system (a) and (b) *x-y* plane (c) and (d) *y-z* plane and synchronization manifold (e) $x_1$-$x_2$ in CS (f) $y_1$-$y_2$ in AS and (g) $z_1$-$z_2$ in CS.

## VI. Discussions and Conclusions

Sometimes counter-rotation may be obtained by replacing any state variables of a system by same state variable with negative sing. But these techniques are not in general and it is system dependent. In this type of coupled counter rotating oscillators there is an uncertainty to get mixed-synchronization. To our best of knowledge, whatever we observed, in some cases there is a possibility to get MS but not for all the cases. As an example, for Rössler system [25] we can create a counter-rotating oscillator by replacing $x$ by $-x$ but MS state is not observed by simple diffusive coupling via $x$ variable or $z$ variable, but it is possible only for y variable. For Lorenz system [26], counter-rotating oscillator is possible by replacement of $z$ by $-z$ but it is not possible by replacement of $y$ by$-y$. If we coupled two counter-rotating Lorenz system via $x$ or $z$ variable, MS scenario is not observed, but MS scenario is possible only for coupling via $y$ variable. So create a counter rotating oscillator is possible by replacing a state variable by its negative sign but MS state is not always exist in that couple counter-rotating oscillators. Although if MS state is exist in coupled oscillators by replacement method, which state variable will be CS or AS is not identified. But in our proposed method we always create counter rotation of a system by changing the conjugate pair of elements (if exist) of the linear matrix and MS state is possible by simple diffusive coupled through any of the variable in the plane of rotation. The type of MS state (which state variable will be CS or AS) is possible to identify using the coupling type.

In summary, we extent work of mixed-synchronization in counter-rotating generalize oscillators where two pair of conjugate elements exist in the linear part of the dynamical systems. We elaborated MS scenario numerically using two chaotic oscillators, Sprott and Pikovsky-Rabinovich model. Noise-induced mixed-synchronization is also elaborated using PR model. Experimental evidence of the extended work is also shown using electronic circuits. Good qualitative agreement is obtained between the numerical and experimental results.

**Acknowledgement**: The authors are very thankful to Syamal K. Dana for helpful discussions and suggestions. Authors are very thankful to referees for making valuable comments which made this paper at the present form possible.